\documentclass{ifacconf}

\usepackage{graphicx}      % include this line if your document contains figures
\usepackage{natbib}        % required for bibliography
\usepackage{amsmath}

\begin{document}
\begin{frontmatter}

\title{Identification of PK-PD Insulin Models using Experimental GIR Data} 

% Johns input: Identification of PK/PD insulin models

\author[DTU,NOVO]{Kirstine Sylvest Freil},
\author[DTU,NOVO]{Liv Olivia Fritzen},
\author[NOVO]{Dimitri Boiroux},
\author[NOVO]{Tinna B. Aradottir},
%\author[NOVO]{??Jeppe, Henrik??},
\author[DTU]{John Bagterp Jørgensen}

\address[DTU]{Department of Applied Mathematics and Computer Science, Technical University of Denmark, DK-2800 Kgs. Lyngby, Denmark}

\address[NOVO]{Novo Nordisk A/S, DK-2860, Søborg, Denmark}

\begin{abstract}  % Abstract of 50--100 words
We present a method to estimate parameters in pharmacokinetic (PK) and pharmacodynamic (PD) models for glucose insulin dynamics in humans. The method combines 1) experimental glucose infusion rate (GIR) data from glucose clamp studies and 2) a PK-PD model to estimate parameters such that the model fits the data. Assuming that the glucose clamp is perfect, we do not need to know the details of the controller in the clamp, and the GIR can be computed directly from the PK-PD model. To illustrate the procedure, we use the glucoregulatory model developed by Hovorka and modify it to have a smooth non-negative  endogeneous glucose production (EGP) term. We estimate PK-PD parameters for rapid-acting insulin analogs (Fiasp and NovoRapid). We use these PK-PD parameters to illustrate GIR for insulin analogs with 30\% and 50\% faster absorption time than currently available rapid-acting insulin analogs. We discuss the role of system identification using GIR data from glucose clamp studies and how such identified models can be used in automated insulin dosing (AID) systems with ultra rapid-acting insulin. 

%In this paper, we propose a method to identify and fit Pharmacokinetic (PK) and Pharmacodynamic (PD) models based on experimental Glucose Infusion Rate (GIR) data to use for simulation of insulin efficiency. By modifying the EGP contribution and rearranging the glucoregulatory model developed by Hovorka, we use experimental PK/PD insulin data from clamp studies to perform parameter estimation and fit the model. This adapted model is then used to simulate insulin analogues with 30\% and 50\% faster absorption time than currently available rapid-acting insulin. Finally, the benefit of this model identification method is considered followed by a discussion on the use of ultra rapid-acting insulin for treatment of insulin dependent diabetes.
\end{abstract}

\end{frontmatter}
%===============================================================================
\section{Introduction}
% Vinklen: Praktisk måde/metode at få modeller, der er afstemt til eksperimental data --> der evt. kan bruges til at simulere insulin (hurtig virkende insulin)

% Figurene skal være EPS format 'EPS to PDF' i Matlab
%Hvorfor er der behov for specifikke modeller? Hvorfor er der behov for bedre fit til eksperimentielt data?

Mathematical simulation models are widely used in the development of new treatments for type 1 diabetes (T1D) and type 2 diabetes (T2D) related to subcutaneous insulin delivery with pump-based automated insulin dosing (AID) systems and pen-based decision support systems. These models are typically population based with parameters obtained to give qualitative reasonable results. The models are rarely fitted to match novel rapid-acting insulin analogs nor ultra rapid-acting insulin analogs in development. Accurate models that fit the experimental data become increasingly important in the development of AID systems as the dosing algorithm is closely related to the PK-PD properties of the insulin analog.
Traditionally, system identification would suggest to obtain the glucose response to insulin injection by a pulse or step test around a given steady state. Such test are not possible with insulin, as any reasonable amount would cause severe hypoglycemia, be very dangerous, and possibly kill the patient. Instead, an approximation to the glucose impulse response can be obtained by the glucose infusion rate (GIR) in a glucose clamp study. This corresponds to closed-loop system identification. GIR curves from glucose clamps are a common practice in the pharmaceutical industry for reporting the effectiveness of a new insulin analog. Consequently, GIR data is publicly available for most insulin analogs \citep[e.g.][]{Haahr:etal:cp2020}. However, these GIR curves have hardly been used for glucose-insulin simulation models.  Furthermore, one can argue that in the future it should be good medical practice to supply such data along with the registration of the drug to enable simulation based therapy evaluation and development.

%

%drug development industry to estimate physiological responses and simulate PK/PD properties e.g. drug absorption, distribution and elimination. Additionally, physiological simulations are used for in silico experiments or to assist in designing actual clinical trials. Thus, it is always of interest to improve models to gain more precise and reliable simulations. A favorable way of doing this, is to fit the models to experimental data. The aim is for the model and its parameters to align with actual data to make the simulations more accurate and trustworthy. 

% Kirstine indsæt kilde til dette

%Drug modeling and simulation are very relevant in the development and administration of artificial insulin in diabetes treatment. Insulin therapy is highly effected by the insulin absorption time and other PK/PD characteristics. One way to investigate insulin efficiency in clinical experiments is to measure plasma insulin concentration and the glucose insulin rate (GIR) in clamp studies. These measures are publicly available for a variety of insulin analogues \citep{Haahr:etal:cp2020}, and it is thus a useful reference for fitting insulin models. 

Euglycaemic glucose clamp studies to obtain GIR curves and estimate the efficiency of insulin analogs is a well established method \citep{Heise:etal:DOM2016}. \cite{Pavan:etal:cmpb2022} describe how euglycemic glucose clamps can be used to estimate the effectiveness of insulin analogs and provide software based on PID algorithms to conduct such glucose clamp studies. They also tested their software using the UVA/Padova T1D simulator \citep{Visentin:etal:jdst2018}. None of these papers used the euglycaemic glucose clamp GIR data to estimate parameters in PK-PD glucose-insulin models. They merely report the GIR curves and describe how to conduct the euglycaemic glucose clamp studies. 

%it can be used to estimat Other research on in silico GIR experiments exists in the literature. \cite{Pavan:etal:cmpb2022} developed a Glucose Clamp Assistant (Gluclas) software based on a Proportional-Integral-Derivative (PID) control algorithm to assist in GIR adjustment. The software was evaluated on the UVA/Padova Type 1 Diabetes Simulator by \cite{Visentin:etal:jdst2018}. Another widely used model for diabetes simulation is the model developed by \cite{Hovorka_2004}. Our work is based on a rearrangement of the Hovorka model.
% Vi skal nok have lidt mere other research ind her

In this paper, we assume that the euglycaemic glucose clamp is ideal (perfect control). This implies that we do not need to know the controller nor the algorithm that was used to obtain GIR data for a subcutaneous insulin bolus impulse and fixed basal insulin injection at its steady state level. We can rearrange the PK-PD model such that the GIR is an output from the model and use that in parameter estimation algorithms for the PK-PD model. We use the PK-PD model by \cite{Hovorka_2004} modified with a smooth non-negative endogeneous glucose production (EGP) function. The use of euglycaemic glucose clamp GIR data for estimation of parameters in PK-PD models is the key novelty in the paper. It is significant because it allows estimation of realistic glucose-insulin PK-PD models from readily available data. These models can be used in the development of pump-based AID for improved diabetes treatment, e.g. by using ultra fast-acting insulin analogs. To indicate this we provide GIR curves for ultra-rapid acting insulin analogs. 

%With this work, we propose a method for fitting insulin- and glucoregulatory models to experimental insulin and GIR data. We describe a rearrangement and modification of the Hovorka model. Then we solve for the required glucose infusion to keep the simulated plasma glucose concentration unchanged at a desired level. Afterwards we perform parameter estimation when fitting for both the plasma insulin concentration (PK part) and the GIR (PD part). This fitted model is then used to simulate insulin analogues with shorter insulin absorption time, which could be used in the research of enhanced insulin efficiency. By using more specific models to simulate insulin and patient response, more accurate and justified decisions can also be made regarding insulin administration e.g. in closed loop treatment systems. The identification method in this paper could also be relevant in the research of other PK/PD drugs and for more accurate in silico simulations.

The remaining sections in this paper are organized as follows. 
Section \ref{sec:ClampStudies} describes the experimental data obtained from clamp studies. In Section \ref{sec:Model}, the modified Hovorka model is rearranged to simulate the GIR. Section \ref{sec:ParameterEstimation} describes the parameter estimation and the obtained fits to PK and PD data. Section \ref{sec:Discussion} discusses the results together with the use of ultra rapid-acting insulin. Finally, Section \ref{sec:Conclusion} concludes our main findings.

\section{Euglycaemic glucose clamp studies}
\label{sec:ClampStudies}
\cite{Pavan:etal:cmpb2022} and \cite{Heise:etal:DOM2016} describe experimental procedures for euglycaemic glucose clamp studies and their applications in measuring the glucose lowering effectiveness of insulin analogs. 
A clamp study is a method to investigate aspects of the glucose metabolism and insulin efficiency \citep{Pavan:etal:cmpb2022}. The experiments are used to analyze insulin properties in relation to its ability to decrease the blood glucose concentration. Various methods for designing clamp studies to investigate insulin effect exist. In clinical clamp studies, a rapid-acting bolus insulin injection is administered to subjects. Subsequently, the blood glucose concentration is maintained ('clamped') at a specific level by carefully adjusting a rate of intravenous glucose infusion. The glucose infusion rate required to maintain the desired blood glucose concentration throughout the experimental time period is measured as the GIR curve. This GIR measurement serves as an indicator of insulin analogs' ability to reduce the blood glucose concentrations and thereby illustrates the insulins' PD properties.  

% TODO - SHORTEN THIS AND WRITE MUCH MORE CONCISESIVELY
The clinical experimental PK-PD data used to fit our simulations are obtained from \cite{Haahr:etal:cp2020}. %research paper: \textit{"Fast‑Acting Insulin Aspart: A Review of its Pharmacokinetic and Pharmacodynamic Properties and the Clinical Consequences"} published at Clinical Pharmacokinetics. 
\cite{Haahr:etal:cp2020} compare the effect of NovoRapid and Fiasp which are rapid-acting insulin analogs. 
%The clinical trials are selected through diverse criteria with one being single-center trials, ensuring consistency in design and procedural approach.
They present PK-PD data for both T1D and T2D patients. In the identification, we focus only on the experiments performed for T1D patients, where the clamp target is 5.5 mmol/L. The PK-PD insulin efficiency data are extracted using version 4.6 of the open source web-based digitizer called 'WebPlotDigitizer' \citep{WebPlotDigitizer}. The program extracts the numerical curve coordinates by applying image extraction algorithms. The extracted data points are afterwards processed in MATLAB.

\begin{figure}[tb]
\begin{center}
\includegraphics[width=0.48\textwidth]{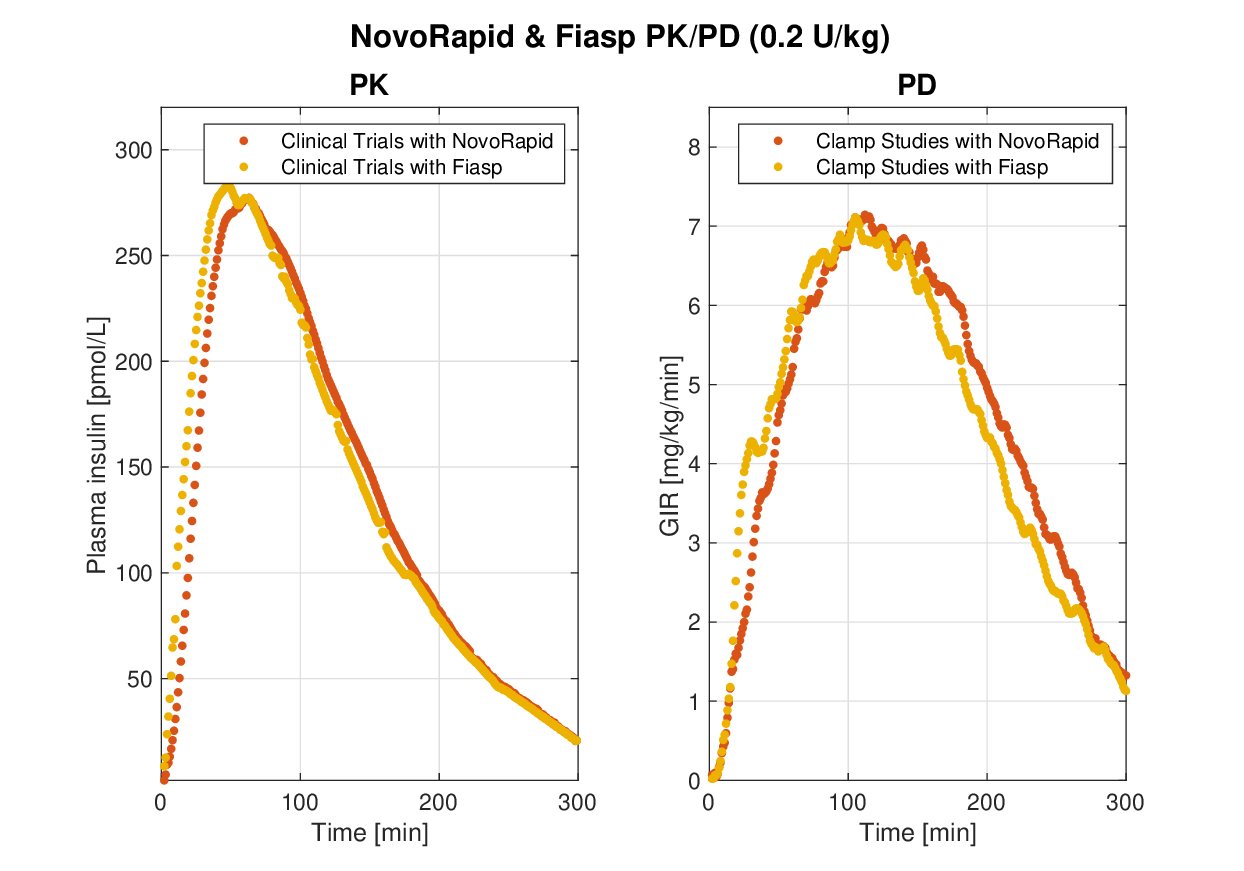}    
\caption{Digitalized PK-PD data for NovoRapid and Fiasp. The data is extracted from \cite{Haahr:etal:cp2020} by the use of a web based plot digitizer \citep{WebPlotDigitizer}.} 
\label{fig:Extracted_PKPD_Data}
\end{center}
\end{figure}

\section{The Model}
\label{sec:Model}

In this paper, we consider a mathematical model represented by an ODE system with initial conditions (Initial Value Problem) that can be written in the form
\begin{subequations}
\begin{align}
    x(t_0) &= x_0 \\
    \dot{x}(t) &= f(t, x(t), u(t), d(t), p)
\end{align}
\end{subequations}
$t_0$ is the initial time, $x_0$ is the initial state, $t$ is time,  $x$ represents the state variables, $u$ is the manipulated input (i.e. administered insulin), $d$ represents disturbances (e.g. meals), and $p$ are the model parameters.

\begin{figure}[tb]
\begin{center}
\includegraphics[width=0.48\textwidth]
{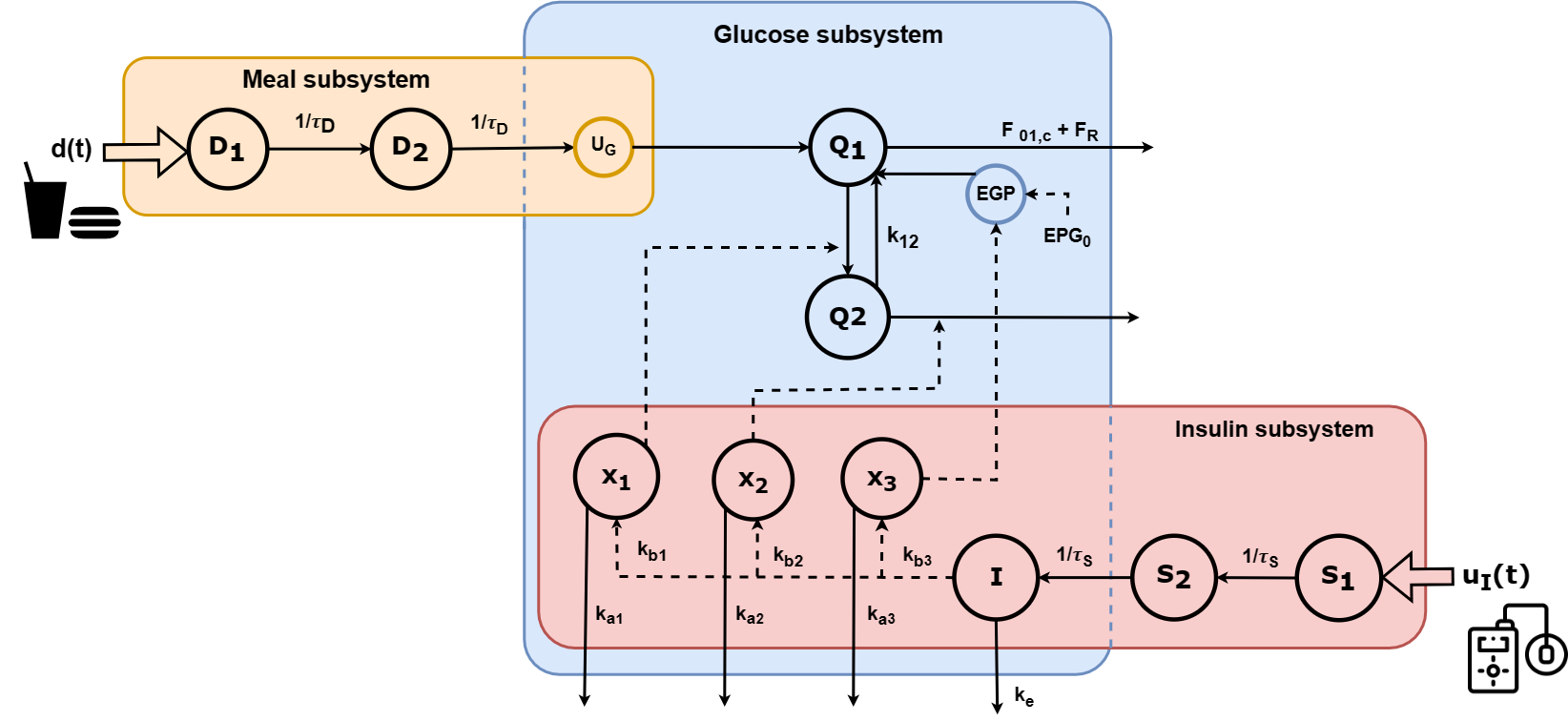}    
\caption{Diagram of the Hovorka model with the meal subsystem, the insulin subsystem, and the glucose subsystem.} 
\label{fig:HovorkaDiagram}
\end{center}
\end{figure}

We use the physiological model referred to as the Hovorka model \citep{Hovorka_2004}. Figure \ref{fig:HovorkaDiagram} is a diagram of the model structure. This gluco-regulatory compartment model describes the relation between meal ingestion, injected insulin, and blood glucose concentration. In this work, the meal subsystem is not relevant since we simulate intravenous glucose injection. %i.e. glucose being injected directly into the blood.

\subsection{Insulin Subsystem}

The insulin absorption subsystem with subcutaneous insulin state variables $S_1$ [mU] and $S_2$ [mU] is modeled by the equations 
\begin{subequations}
\begin{align}
    \dot{S}_1 (t) &= u_I(t) - \frac{S_1 (t)}{\tau_S} \\
    \dot{S}_2 (t) &= \frac{S_1 (t)}{\tau_S} - \frac{S_2 (t)}{\tau_S}
\end{align} 
\end{subequations}
 $u_I$ [mU/min] is the insulin administration rate and $\tau_S$ [min] represents the insulin absorption time constant. 
The insulin appears in the plasma at the rate $S_2(t)/\tau_S$.
The insulin concentration in plasma, $I(t)$ [mU/L], is 
\begin{equation}
    \dot{I}(t) =   \left( \frac{S_2(t)}{\tau_S} \right) \frac{1}{V_I}  - k_e I
\end{equation}
$V_I$ [L] is the volume in which the incoming insulin is distributed and $k_e$ [1/min] is the insulin elimination rate. 

The insulin actions on glucose kinetics are modelled as a three-compartment model. Here $x_1$ [1/min] represents the insulin influence on transport and distribution, $x_2$ [1/min] covers the influence on the glucose in adipose tissue, and $x_3$ [1/min] denotes the influence on the endogenous glucose production in the liver \citep{Boiroux:etal:DYCOPS2010}.
\begin{subequations}
\begin{align}
\dot{x}_1(t) &= k_{b1}I(t)-k_{a1}x_1(t) \\
\dot{x}_2(t) &= k_{b2}I(t)-k_{a2}x_2(t) \\
\dot{x}_3(t) &= k_{b3}I(t)-k_{a3}x_3(t)
\end{align}
\end{subequations}
The parameters $k_{ai}, \; i=1, 2, 3$ are the deactivation rates, and the parameters $k_{bi}, \;i=1, 2, 3$ are the activation rates. 

\subsection{Glucose Subsystem}
The glucose subsystem is divided into the accessible glucose in the plasma, $Q_1$ [mmol], and the non-accessible glucose in the muscle and adipose tissue, $Q_2$ [mmol]. These are modeled as
\begin{subequations}
    \begin{align}
     \dot{Q}_1(t) &= U_G  - F_{01,c} - F_R -R_{12} + EGP \\
    \dot{Q}_2(t) &= R_{12} - R_2
\end{align}
\end{subequations}
with the inter compartment transport rate equations
\begin{subequations}
\begin{align}
 R_{12} &= Q_1(t)x_1(t)-k_{12}Q_2(t) \\
R_2 &= Q_2(t)x_2(t)   
\end{align}
\end{subequations}
$U_G$ is the incoming glucose rate [mmol/min], $R_{12}$ is the glucose transferred from $Q_1$ to $Q_2$, $R_2$ is the glucose transfer out of $Q_2$, and $k_{12}$ [1/min] is the transfer rate between compartments.

The endogenous glucose production, $EGP$ [mmol/min], from the liver is described as
\begin{equation}
    EGP = EGP_0(1-x_3(t))
\end{equation}
$EGP_0$ [mmol] is the endogenous glucose production extrapolated to zero insulin concentration. 

%The terms $F_{01,c}$ and $F_R$ depend on the value of 
The plasma glucose concentration, $G$ [mmol/L], is defined as
\begin{equation}
    G(t) = \frac{Q_1(t)}{V_G}
\end{equation}
where \textit{$V_G$} $[L]$ is the glucose distribution volume.

The term $F_{01,c}$ represents the insulin independent glucose flux, and $F_R$ covers the renal glucose clearance. They are described as
\begin{subequations}
    \begin{align}
    F_{01,c} &= 
    \begin{cases}
    F_{01},  &G(t) \geq 4.5 \text{ mmol/L} \\
    F_{01}G(t)/4.5, & \text{otherwise}
        \label{eq:fc}
    \end{cases} \\
     F_R &= 
    \begin{cases}
   0.003 (G(t)-9.0) V_G,  & G(t) \geq 9.0 \text{ mmol/L} \\
    0, & \text{otherwise}
    \end{cases} 
    \label{eq:f01c}
        \end{align}
\end{subequations}
$F_{01}$ [mmol/min] is the nominal total non-insulin dependent glucose flux.

\subsection{Modified EGP}

In the original Hovorka model, the endogenous glucose production (EGP) that results from glycogenolysis and gluco-neogenesis processes occurring in the liver is defined as
\begin{equation}
    EGP = EGP_{0}(1-x_{3}(t))
    \label{eq:EGP_Original}
\end{equation}

During the GIR simulation with the original Hovorka model, we observed negative EGP values, which is contradictory with it being a production. Different modifications are considered to ensure only positive values; the ReLu, the softplus, and the scaled softplus version.  
\begin{subequations}
\begin{align}
     EGP_{ReLU} &= \max(0, EGP)
     \\
     EGP_{softplus} &= \log(1+e^{EGP})
    \\
    EGP_{scaled} &= A + \log(1+e^{(B \cdot C ((1-x_3(t))-D)})/C
\end{align}
\end{subequations}
$A = 0$ is the value after activation, $B=EGP_0$ is the starting point, $C=10/EGP_0$ represent the curve steepness, and $D=0$ is the point of activation.

\begin{figure}
\begin{center}
\includegraphics[width=0.48\textwidth]{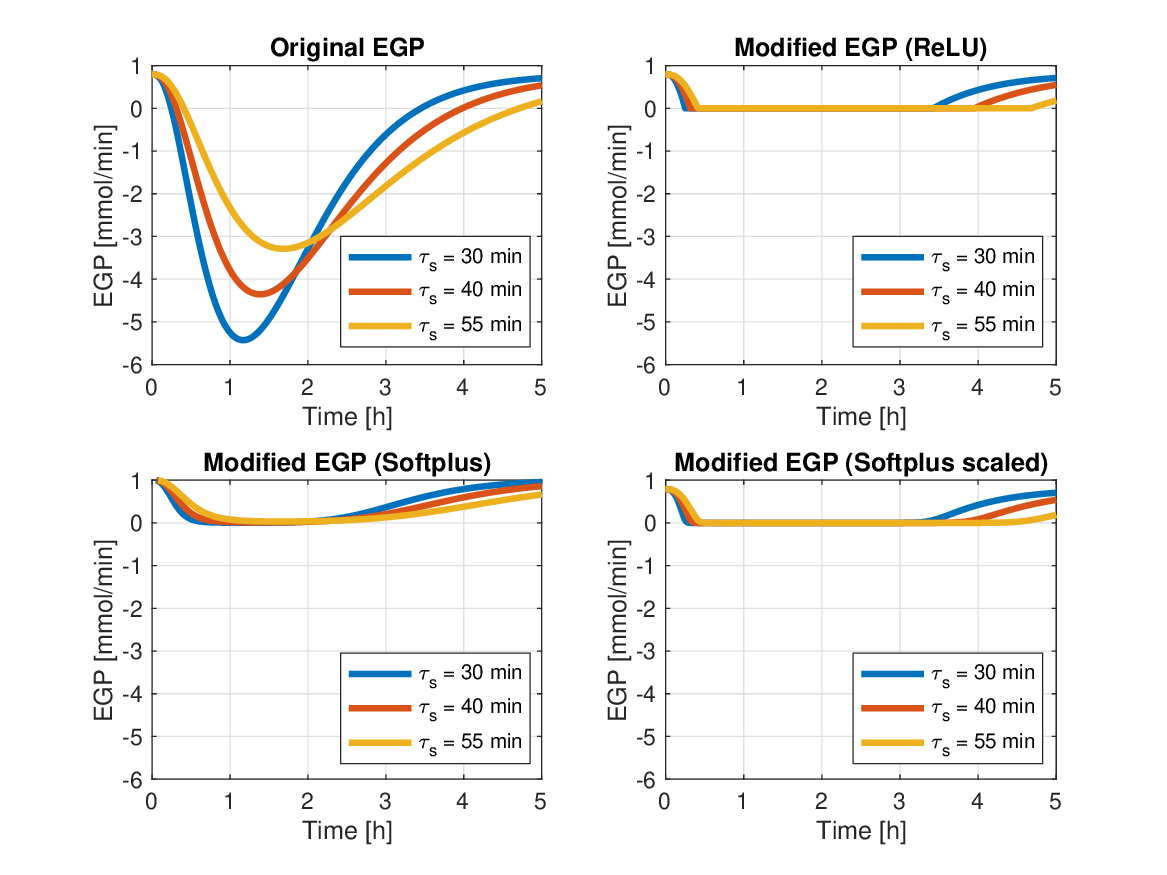}    
\caption{The original EGP  from \cite{Hovorka_2004} compared to the modified versions ReLU, Softplus and Softplus scaled for different insulin absorption's times $\tau_s$ for a defined bolus injection at $14000$ mU and $\bar{G} = 5.5$ mmol/L. } 
\label{fig:EGP_visu}
\end{center}
\end{figure}

Figure \ref{fig:EGP_visu} shows the EGP versions during the GIR simulations plotted for different insulin absorption times. To resemble the ReLu function but with a smooth curve, we decided to apply the scaled softplus version as a modified EGP in this work.

\subsection{GIR Rearrangement}
The Hovorka model is rearranged to represent the mechanism in a GIR experiment, where the blood concentration is held constant at $\bar{G} = 5.5$  mmol/L. Accordingly, the change in blood glucose concentration, $\dot{Q}_1$, is equal to 0. 
\begin{equation}
     \dot{Q}_1(t) = 0
\end{equation} 
Consequently, the accessible plasma glucose is
\begin{equation}
    Q_1(t) = V_g \bar{G}(t) = V_g \cdot 5.5 \; \text{mmol/L}
\end{equation}

\begin{figure}[tb]
\begin{center}
    \centering
    \includegraphics[width=0.46\textwidth]{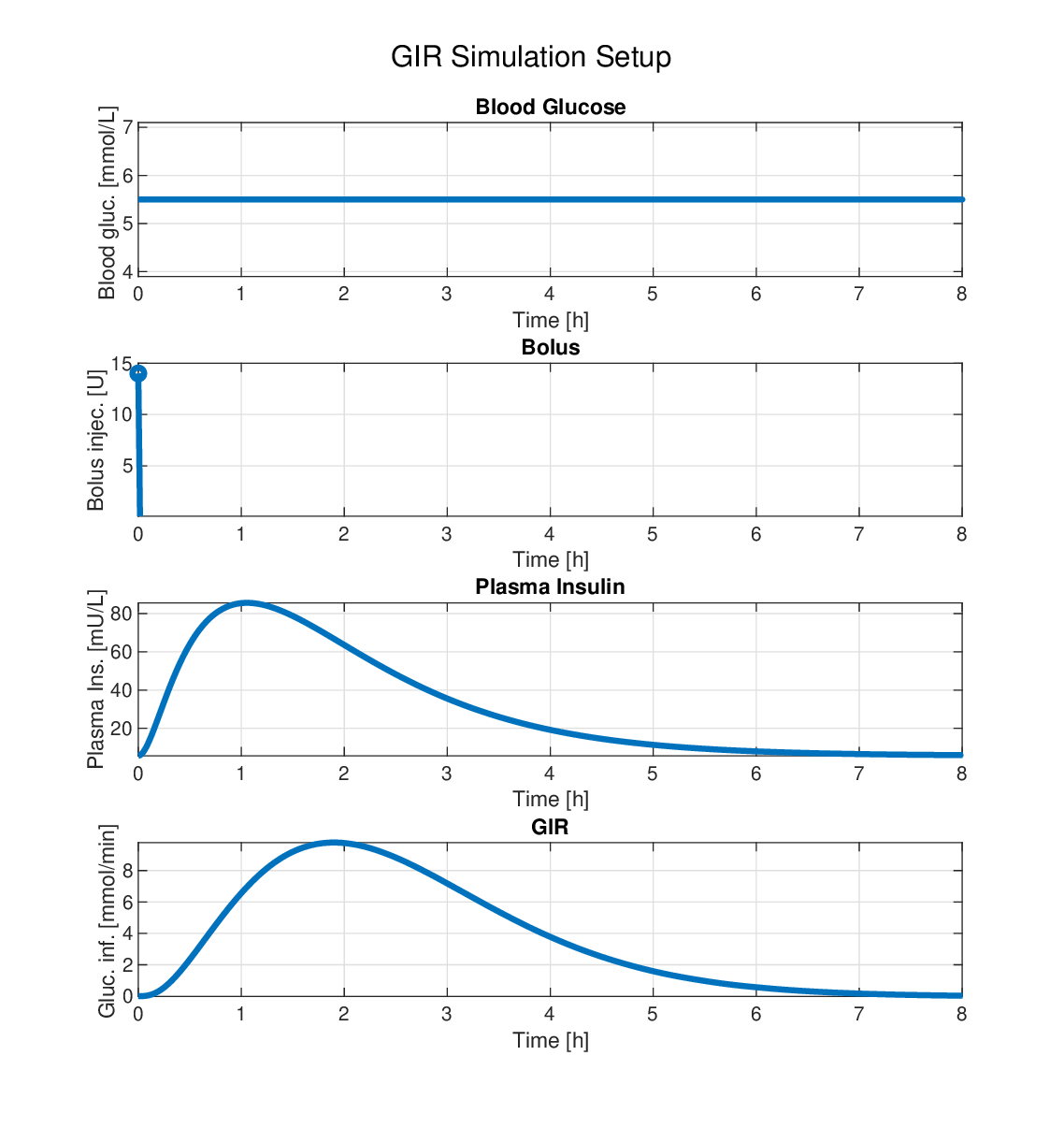}
    \caption{GIR simulation setup with a blood glucose concentration clamped at $\bar{G}=5.5$ mmol/L, a bolus insulin injection of 14 U, and the resulting plasma insulin concentration ($I(t)-I_{ss}$), and the required GIR ($U_G(t)$).}
    \label{fig:GIR_setup}
\end{center}
\end{figure}

 Figure \ref{fig:GIR_setup} shows the GIR simulation setup. To resemble the published data, we used a bolus of $0.2$ U/kg \citep{Walpole:etal:BMC2012}. Hence, the  bolus given is $14$ U  for an average European person at 70 kg. This setup is used to simulate incoming plasma insulin concentration from the bolus injection, and the required glucose, $U_G$ (the GIR curve). The response to this bolus is simulated for 8 hours for the constant glucose concentration and a corresponding basal insulin rate, $u_I$. %corresponding to the steady state glucose concentration. 

The steady states values ($_{ss}$) for a blood glucose concentration of 5.5 mmol/L are used as starting conditions in the simulation, including the insulin bolus injection added as an impulse to $S_1$. 
\begin{equation}
    x_0 = \begin{bmatrix}
        Q_{2,ss} \\
        S_{1,ss} \\
        S_{2,ss} \\
        I_{ss} \\
        x_{1,ss} \\
        x_{2,ss} \\
        x_{3,ss} \\
    \end{bmatrix}
    + 
    \begin{bmatrix}
        0 \\
        \text{bolus} \\
        0 \\
        0\\
        0\\
        0\\
        0
    \end{bmatrix}
\end{equation}
The resulting states throughout this simulation are used to compute the plasma insulin concentration, $I(t)$ (PK), and compute the required glucose infusion rate, $U_G(t)$ (PD). The key trick is to recognize that assuming perfect glucose control
\begin{equation}
    \dot{Q}_1(t) = 
     U_G - F_{01,c} - F_R - R_{12} + EGP = 0,
\end{equation}     
provides an expression for the GIR 
\begin{equation}
    U_G = F_{01,c} + F_R + R_{12} - EGP
\end{equation}

\section{Parameter Estimation}
\label{sec:ParameterEstimation}

We use parameter estimation to fit the modified Hovorka model to the experimental data. For simplicity, we start by fitting the PK part by determining the three parameters in the insulin subsystem; the insulin absorption time, $\tau_S$, the insulin elimination rate, $k_e$, and the insulin distribution volume, $V_I$. Using the fitted PK parameters, we subsequently fit the PD part (the GIR curves), by estimating the insulin sensitivity related parameters, $k_{b1}$,  $k_{b2}$,  $k_{b3}$.

\subsection{PK Parameter Estimation}

\begin{figure}[tb]
\begin{center}
\includegraphics[width=0.46\textwidth]{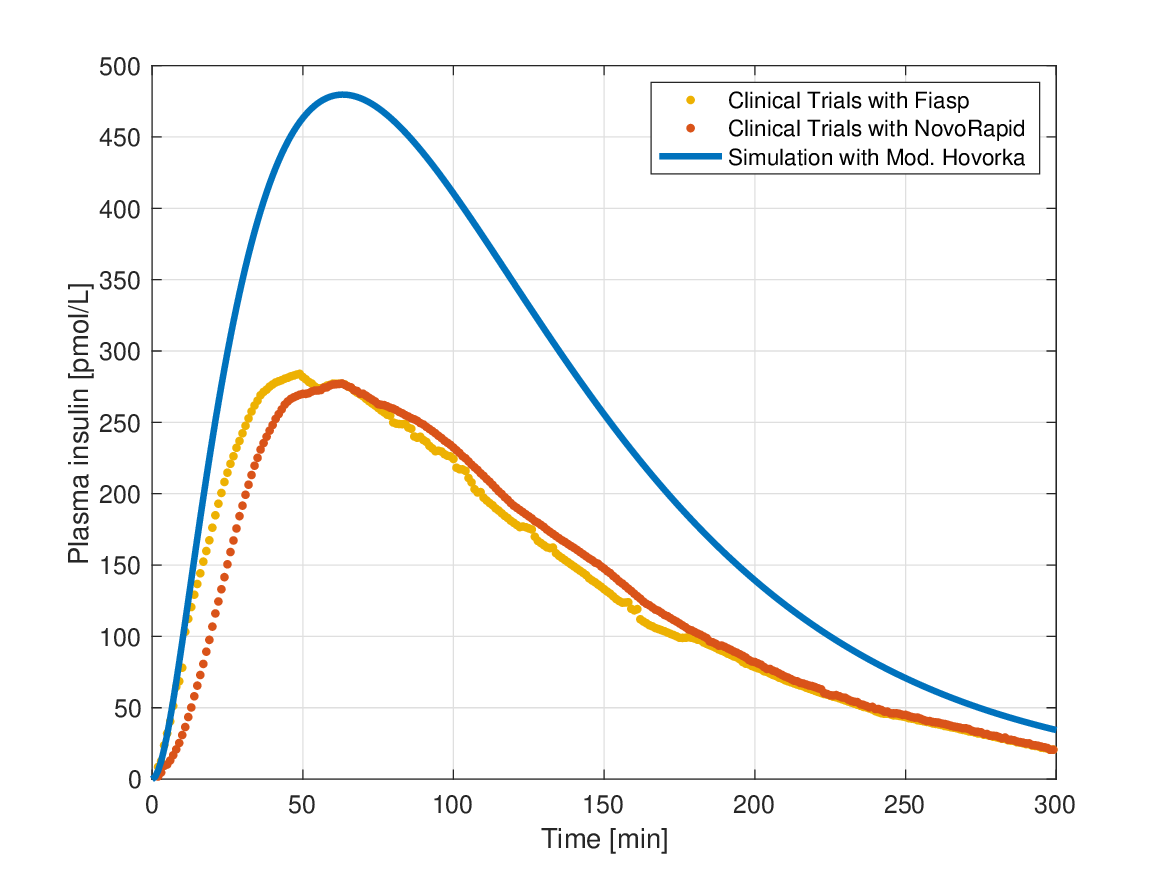}    
\caption{Experimental plasma insulin ($I-I_{ss}$) concentration curves for Fiasp and Novorapid from clamp studies together with the simulated modified Hovorka model curve prior to parameter estimation.} 
\label{fig:PI_Before_PE}
\end{center}
\end{figure}

Figure \ref{fig:PI_Before_PE} shows a comparison of the simulated plasma insulin concentration ($I-I_{ss}$) with the original parameter values plotted together with the experimental plasma insulin data for both NovoRapid and Fiasp. By inspection, it is obvious that the modified Hovorka model does not fit the data well.  

The plasma insulin plot is the insulin response to that specific bolus injection, and thus does not include the share of insulin from any prior or ongoing basal injection. Therefore, the steady state value of plasma insulin is subtracted from the fitted simulation. The basal insulin is required to make the model start in steady state with a blood glucose of $\bar{G} = 5.5$  mmol/L and end in steady when the bolus injection is eliminated from the plasma. % is subtracted from the fitted simulation.

\begin{figure}[tb]
\begin{center}
\includegraphics[width=0.48\textwidth]
{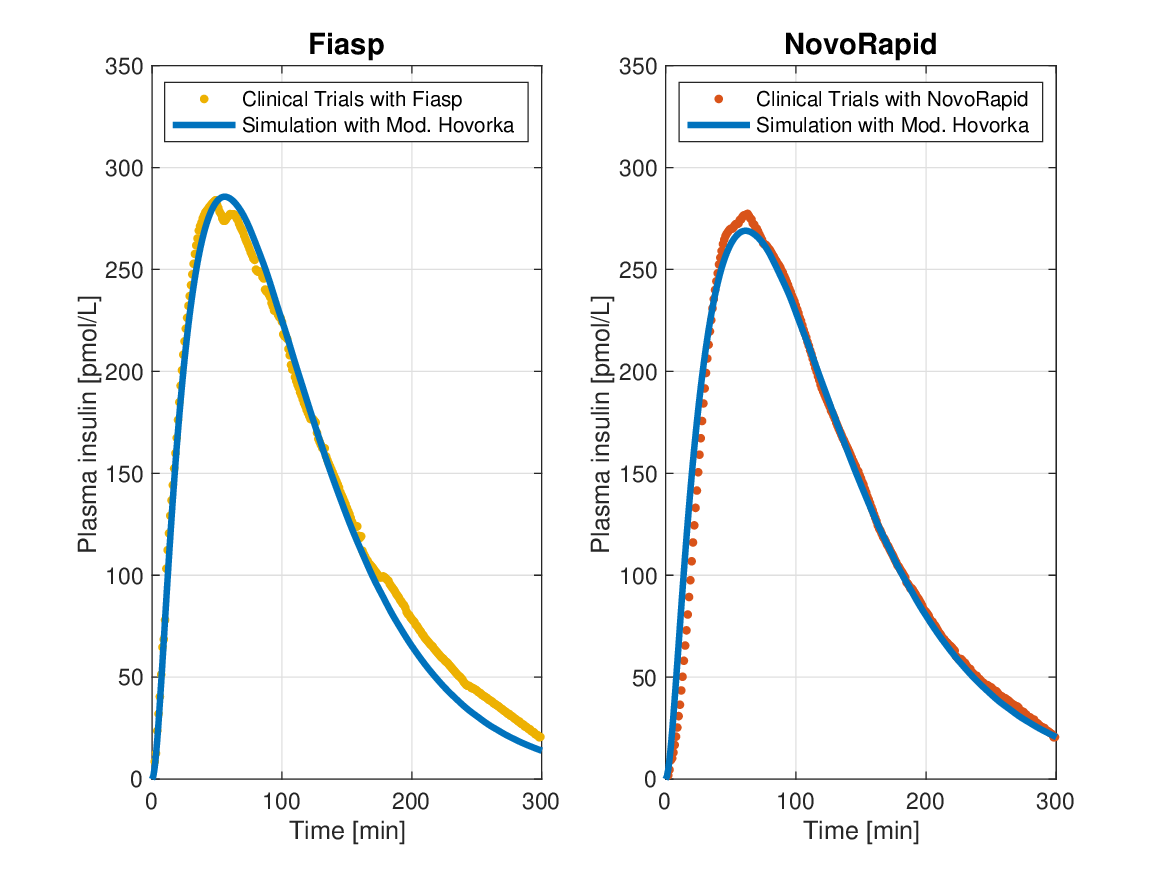}    
\caption{Plasma insulin ($I-I_{ss})$ curves for experimental and simulated Fiasp and NovoRapid insulin data posterior to parameter estimation.} 
\label{fig:PI_after_PE}
\end{center}
\end{figure}

The cost function is defined as the sum of squared error (SSE) between the actual experimental data and the simulated data for both Fiasp and NovoRapid.
\begin{equation}
\begin{split}
J_{PK}(\theta_{PK})
 & = \sum_{i=1}^n (I_{actual,FI}-I_{sim,FI})^2 \\
 & + \sum_{i=1}^n (I_{actual,NR}-I_{sim,NR})^2 \\
& \theta_{PK} = \{\tau_{S,FI}, \tau_{S,NR}, k_{e,FI}, k_{e,NR}, V_I\}
 \end{split}
\end{equation}
 The Matlab function {\tt fmincon} is used to minimize the cost function in a bound constrained nonlinear least squares problem  where the fitted parameters must be in an interval determined by  upper and lower bounds. 
\begin{equation}
    \begin{aligned}
      \min_{\theta_{PK}} \quad & J_{PK}(\theta_{PK})  \\
      s.t. \quad & \theta_{PK,lb} \leq \theta_{PK} \leq \theta_{PK,ub}
\end{aligned}
\end{equation}
Along with experimental PK data for NovoRapid and Fiasp, Figure \ref{fig:PI_after_PE} displays the simulated PK curves for parameters obtained by the PK parameter estimation. The fit is very good.  

\subsection{PD Parameter Estimation}

The estimated PK values are then used to fit the GIR curves for Fiasp and NovoRapid, respectively. The GIR curves are fitted by adjusting the insulin sensitivity parameters ($k_{b1}$,  $k_{b2}$,  $k_{b3}$) to minimize the cost function containing the sum of squares between the actual (measured) and simulated GIR values (SSE).
\begin{equation}
\begin{split}
J_{PD}(\theta_{PD})
 & = \sum_{i=1}^n (U_{G,actual,FI}-U_{G,sim,FI})^2 \\
 & + \sum_{i=1}^n (U_{G,actual,NR}-U_{G,sim,NR})^2 \\
 & \theta_{PD} = \{k_{b1},k_{b2}, k_{b3}\}
 \end{split}
\end{equation}
Rather than fitting individual PD models, we fit a single PD model for NovoRapid and Fiasp, since the PD mechanism of these insulin analogs is the same.
The cost function is minimized subject to upper and lower bounds for the parameters as well as the constraint that the GIR ($U_G$) should only attain positive values, since the infusion cannot be negative. 
\begin{equation}
    \begin{aligned}
      \min_{\theta_{PD}} \quad & J_{PD}(\theta_{PD}) \\
      s.t. \quad & \theta_{PD,lb} \leq \theta_{PD} \leq \theta_{PD,ub} \\
      & U_G \geq 0
\end{aligned}
\end{equation}

\begin{figure}[tb]
\begin{center}
\includegraphics[width=0.48\textwidth]{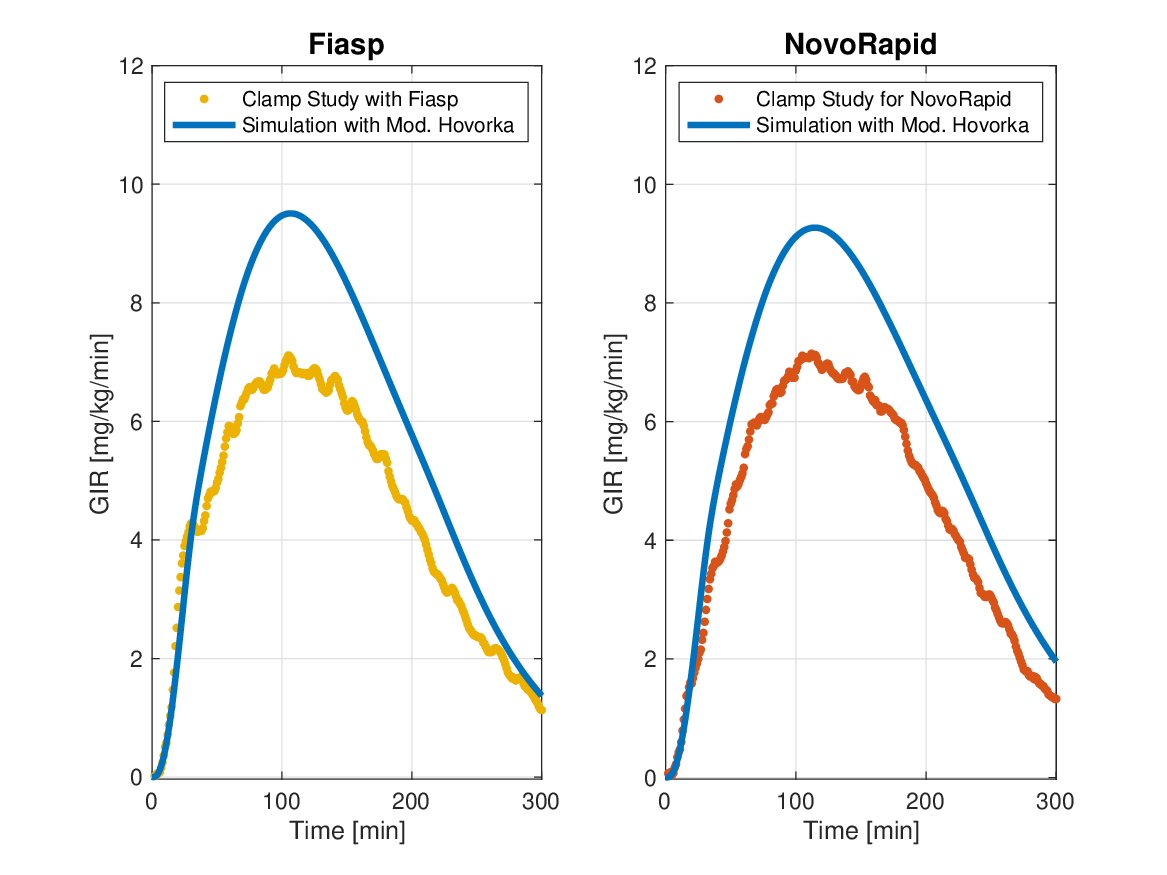}    
\caption{Experimental and simulated GIR curves for Fiasp and NovoRapid prior to the GIR parameter estimation.} 
\label{fig:GIR_Before_PE}
\end{center}
\end{figure}

\begin{figure}[tb]
\begin{center}
\includegraphics[width=0.48\textwidth]{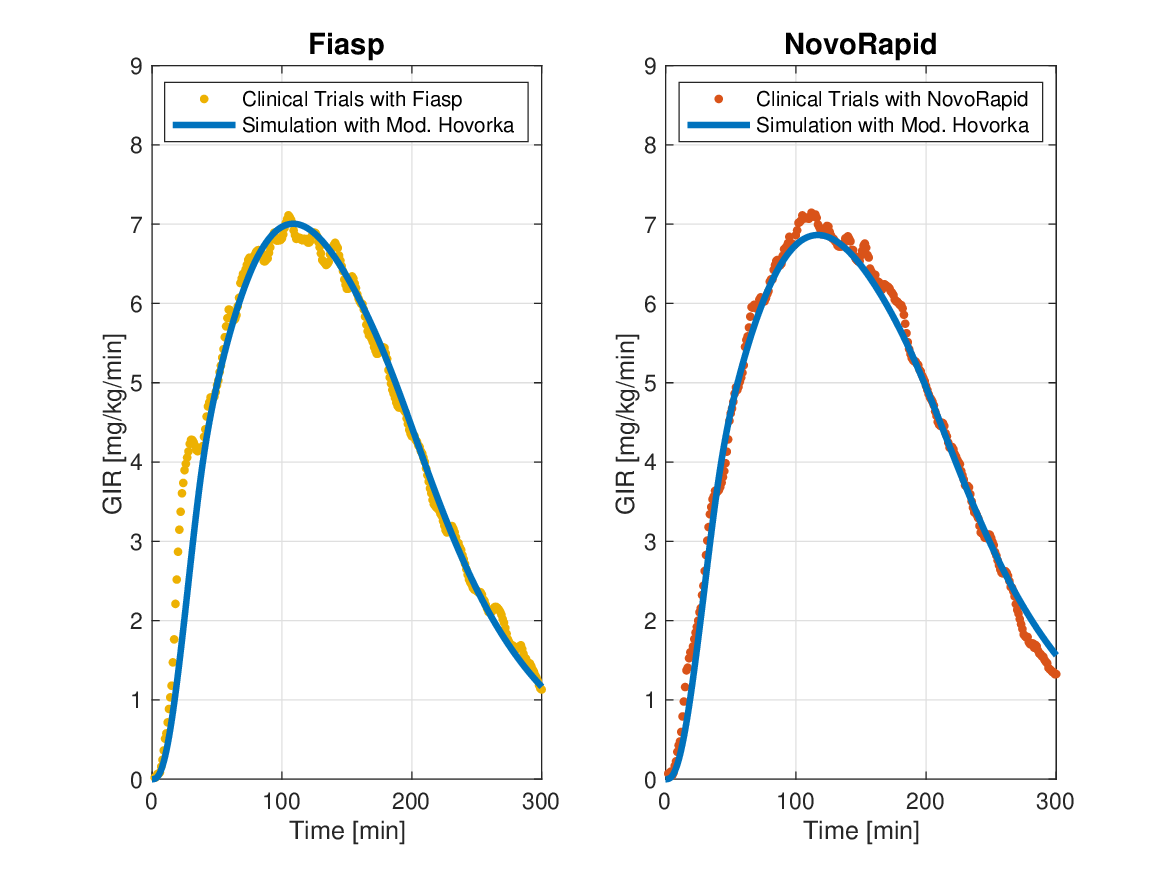}    
\caption{GIR curves for experimental and simulated Fiasp and NovoRapid insulin data with $\bar{G} = 5.5$ mmol/L posterior to GIR parameter estimations.} 
\label{fig:GIR_After_PE}
\end{center}
\end{figure}

%\section{Simulation Results}
%\label{sec:SimulationResults}
Figure \ref{fig:GIR_Before_PE} shows the simulated GIR curves prior to PD parameter estimation along with the experimental PD data for both NovoRapid and Fiasp. Figure \ref{fig:GIR_After_PE} displays the simulated PD curves with the insulin sensitivity parameters ($k_{b1}$, $k_{b2}$, $k_{b3}$) obtained by PD parameter estimation. Table \ref{tab:parameters} provides the standard model parameter values as well as the parameter values estimated from the euglycemic clamp study.

\begin{table}[tb]
    \centering
    \caption{Parameters for the model by \cite{Hovorka_2004}, the Fiasp estimation, and the NovoRapid (NR) estimation.}
    \begin{tabular}{ccccc}
    \hline
     &  & Hovorka & Fiasp & NR\\
    \hline
     $BW$ & kg  & 70 &  \\
     $EGP_0$ & mmol/min  & $0.0161 \cdot BW$ &  \\
     $F_{01}$ & mmol/min & $0.0097 \cdot BW$ &  \\
     $k_{12}$ & 1/min & 0.066  &  \\
     $k_{a1}$ & 1/min & 0.006  & \\
     $k_{a2}$ & 1/min  & 0.06  & \\
     $k_{a3}$ & 1/min & 0.03  & \\
     $V_G$ & L & $0.16 \cdot BW$ &  \\ \hline
     $k_{b1}$ & (L/mU)/min  & $3.07\cdot 10^{-5}$ & $1.73\cdot 10^{-5}$ \\
     $k_{b2}$ & (L/mU)/min & $4.92\cdot 10^{-5}$  & $5.38\cdot 10^{-5}$ \\
     $k_{b3}$ & (L/mU)/min & 0.0016  & 0.0011 \\
    $V_I$ & L & $0.120 \cdot BW$ & $0.137 \cdot BW$  \\ \hline
    $\tau_S$ & min & 55 & 51.1 & 56.1\\
    $k_e$  & 1/min & 0.138 & 0.220 & 0.212 \\
    \hline
    \end{tabular}
    \label{tab:parameters}
\end{table}

\section{Discussion}
\label{sec:Discussion}

% Discuss simulation results, model performance, and the ultra fast acting insulin in diabetes treatment. When new analogues are developed and GIR experiments are performed, the models can likewise be fitted to them, to determine their PK/PD properties. Pumps and hybrid/fully closed loop systems. 

% Måske diskutere at eliminationsraten ikke nødvendigvis bliver hurtigere?

 Figure \ref{fig:PI_after_PE} and Figure \ref{fig:GIR_After_PE} display the simulated modified Hovorka model that is fitted to the available experimental PK-PD data for the rapid-acting insulin analogs NovoRapid and Fiasp. Visual inspection of the results of the method for identifying insulin models shows a high level of accuracy and ability to represent simulated PK-PD properties. The modified Hovorka model fits the actual data very accurately which makes the simulations reliable. Figures \ref{fig:PI_Before_PE} and \ref{fig:GIR_Before_PE} show that the model with standard parameters does not fit the data well. To have representative simulations, it is important that the model is fitted to data.
 %
 %Table \ref{tab:parameters} shows that $\tau_S$ is almost unaltered while elimination and gains change. %is almost the same for the original Hovorka model and NR. % while the gains (insulin sensitivity) and the elimination rate are different.
% Evt. acceptable/adequate?

Within medical drug development of new insulin therapies, experimental PK-PD data is obtained to investigate its efficiency. Using the methods proposed in this paper,  mathematical models can a be fitted to the actual data for the insulin analogs and  used to obtain more reliable simulations. Simulations of the PK-PD properties for new therapies can be used to design and improve new therapies with either pens or pumps as in e.g. AID systems. Simulations can in addition aid in the design of actual clinical trials to obtain approval in drug development phases. When new insulin analogs are developed, such as ultra rapid-acting insulin analogs, fitted models and thereby reliable simulations of the insulin efficiency can benefit the adjustments of dosing algorithms to ensure safe treatment management.

\begin{figure}[tb]
\begin{center}
\includegraphics[width=0.37\textwidth]{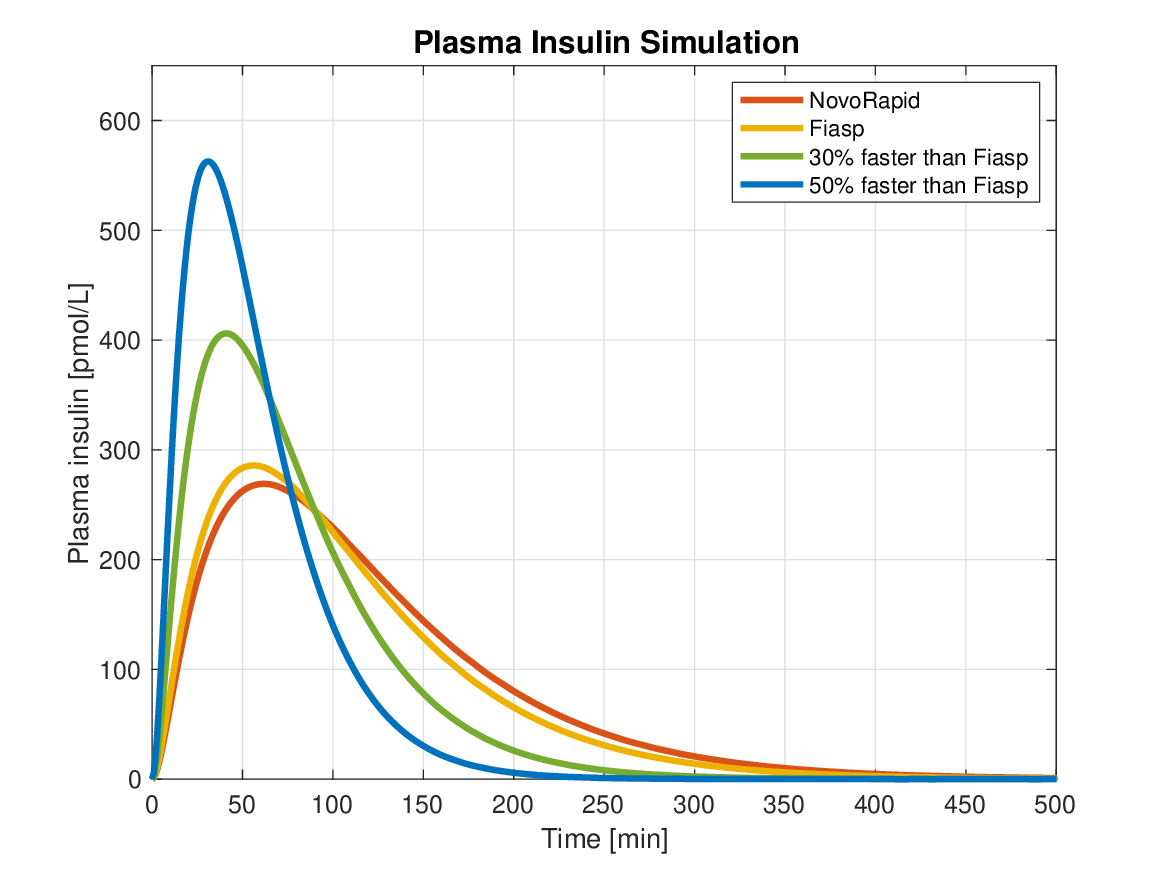}    
\caption{Plasma insulin ($I-I_{ss}$) simulations with the modified Hovorka model for NovoRapid, Fiasp, and insulin analogs with 30\% and 50\% faster absorption time than Fiasp.} 
\label{fig:PI_Simulation}
\end{center}
\end{figure}

\begin{figure}[tb]
\begin{center}
\includegraphics[width=0.37\textwidth]{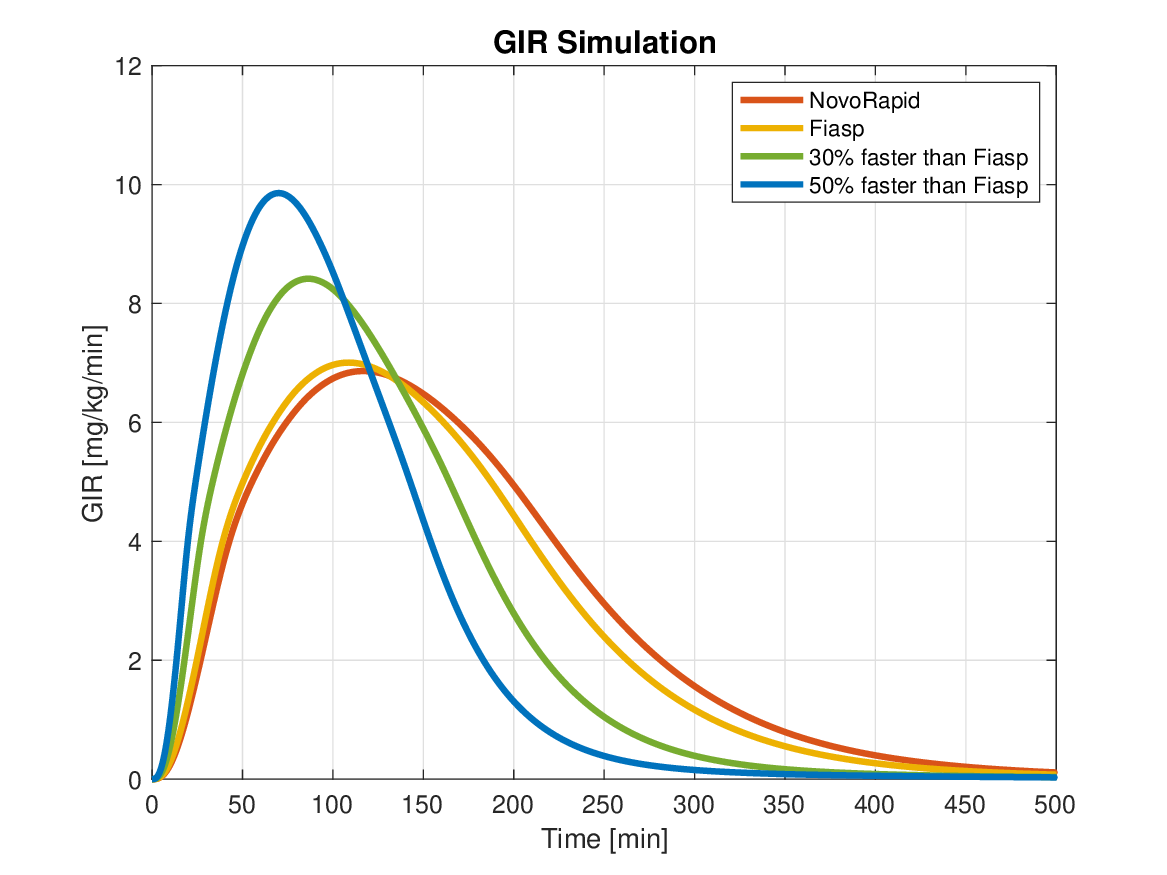}    
\caption{GIR simulations with the modified Hovorka model for rapid-acting insulin NovoRapid, Fiasp, and insulin analogs with 30\% and 50\% faster absorption time than Fiasp.} 
\label{fig:GIR_Simulation}
\end{center}
\end{figure}

Figure \ref{fig:PI_Simulation} shows the plasma insulin simulations for insulin analogs 30\% and 50\% faster than Fiasp along side the fitted simulations for Fiasp and NovoRapid. Figure \ref{fig:GIR_Simulation} displays the corresponding GIR simulations.

Novel ultra rapid-acting insulin analogs faster than what currently is available on the marked, with effects similar to the simulations in Figures \ref{fig:PI_Simulation} and \ref{fig:GIR_Simulation}, can highly improve the development of fully closed loop treatment system for patients with insulin dependent diabetes. In fully closed loop systems, announcements for disturbances are not necessary \citep{Templer:etal:FEL2022}. Disturbances could be meal consumption, exercise or other blood glucose fluctuation factors. Faster novel rapid-acting insulin analogs would be very beneficial 
in future treatment systems without announcements, and thus it would make the managing for patients with diabetes less demanding. Due to the significantly different GIR curves for ultra rapid-acting insulin analogs, the algorithms in AID systems must be redesigned.

\section{Conclusion}
\label{sec:Conclusion}

In this paper, we presented a method for  identification of insulin PK-PD models using experimental GIR data. This approach assumes perfect glucose control such that the model can be rearranged for parameter estimation without needing knowledge about the control algorithm used in the clamp study. The parameter estimation can be conducted using publicly available experimental PK-PD curves. The modified Hovorka model is fitted to experimental GIR data for rapid-acting insulin NovoRapid and Fiasp. The method shows very accurate fits such that these models can be used in development of new insulin analogs. Simulation of novel ultra rapid-acting insulin analogs can help in the development of fully closed loop systems, where the need for meal announcement could potentially be eliminated.

%\begin{ack}
%Place acknowledgments here.
%\end{ack}

\bibliography{ifacconf}             % bib file to produce the bibliography
                                                     % with bibtex (preferred)

                                                                         % in the appendices.
\end{document}